# From closed to open access: A case study of flipped journals

Fakhri Momeni[1], Philipp Mayr[1], Nicholas Fraser [2] and Isabella Peters[3]

*[1] firstname.lastname@gesis.org*
GESIS – Leibniz Institute for the Social Sciences, Unter Sachsenhausen 6-8, 50667 Cologne (Germany)

*[2] N.Fraser@zbw.eu, [3] I.Peters@zbw.eu*
ZBW – Leibniz Information Centre for Economics, Düsternbrooker Weg 120, 24105 Kiel (Germany)

**Abstract**
In recent years, increased stakeholder pressure to transition research to Open Access has led to many journals 'flipping' from a toll access to an open access publishing model. Changing the publishing model can influence the decision of authors to submit their papers to a journal, and increased article accessibility may influence citation behaviour. The aim of this paper is to show changes in the number of published articles and citations after the flipping of a journal. We analysed a set of 171 journals in the Web of Science (WoS) which flipped to open access. In addition to comparing the number of articles, average relative citation (ARC) and normalized impact factor (IF) are applied, respectively, as bibliometric indicators at the article and journal level, to trace the transformation of flipped journals covered. Our results show that flipping mostly has had positive effects on journal's IF. But it has had no obvious citation advantage for the articles. We also see a decline in the number of published articles after flipping. We can conclude that flipping to open access can improve the performance of journals, despite decreasing the tendency of authors to submit their articles and no better citation advantages for articles.

**Introduction**

The term "flip" is often used to describe a journal transitioning from a subscription-based (or closed access (CA)) to an open access (OA) publishing model (Solomon, Laakso & Björk, 2016). This change in the publishing model might have consequences for the journal's future development, for example in terms of citation rates or publishing volumes. Most evidence points towards an open access citation advantage for OA articles over CA articles overall (OACA[1]; Piwowar & Vision, 2013; Sotudeh et al., 2015; Swan, 2010), although the citation advantage is found to be driven primarily by 'Green' and 'Hybrid' OA whilst articles published in 'Gold' OA journals conversely show a disadvantage in terms of citations versus CA articles (Archambault et al., 2014; Piwowar et al., 2018). Swan (2010) reported the results of studies on the OACA across multiple fields and showed significant variability between fields. McKiernan et al. (2016) compared the relative citation rate of OA articles for 19 fields and showed the highest citation advantage for agricultural studies, physics/astronomy and medicine, but a citation disadvantage for ecology and communications studies. A randomized control trial conducted by Davis (2011) could also not confirm the citation advantage of OA journals over CA journals. Hence, there remain many unanswered questions in what drives this erratic behaviour. In addition, there are few studies that systematically research the effects flipping has on a journal's standing, in terms of its citation rates and impact factor (IF) as well as its publication output. One exception is a study by Busch (2014a) who demonstrated an increase in IFs for five journals published by BioMed Central after flipping to an OA model. But, he also found a reduction in the number of articles published by four journals of this group (Busch 2014b).

An important concern for journals is to find an alternative source of income to reader subscription charges when flipping to OA. One alternative source is the author-pays model that shifts the publication costs to authors or their institutions. This may generate financial barriers for some authors who wish to submit their articles to such journals. According to

---

[1] The Open Access Citation Advantage Service (OACA) list [Internet]. Sparceurope.org. 2017 [accessed 27 January 2019]. Available from: https://sparceurope.org/what-we-do/open-access/sparc-europe-open-access-resources/open-access-citation-advantage-service-oaca/

Solomon, Laakso & Björk (2016), transitioning a journal to an OA model for those societies with low numbers of publications can be expensive. Björk (2012) additionally demonstrated that the author-pays model in hybrid journals is unpopular with authors. Peterson et al. (2013) argued that the cost of article processing charges (APC) in this model is often too much for many authors, and publishers try to solve this problem in different ways such as fee waiver policies, subsidizing academic publishing directly without profiteering intermediaries, etc.

Such consequences are of immense importance to those responsible for journal management, thus in-depth, longitudinal bibliometric studies can help to inform decision making of publishers and societies, and their assessment of chances and risks of flipping their journals. A recent study on "reverse-flipped" journals by Matthias et al. (2019) shows that a majority of these journals had an experience of flipping to OA before, pointing out that flipping was not successful for them for different reasons. Moreover, authors need to know what to expect from flipped journals, and whether the move from CA to OA negatively or positively affects the quality of the journal, its reputation, or the chances to get articles published.

This is a work in progress paper presenting the first results of a study on the evolution of bibliometric indicators for a sample of journals, over a period of several years before and after the journal flipped to an OA publishing model. We assess the effect of the journals flipping to OA by studying the changes in the number of published articles, as well as changes in citation metrics at the article (average relative citations) and journal (IF) level[2].

**Data and methods**

A list of journals that have flipped to OA, and the years in which they flipped were retrieved from the Open Access Directory (a wiki where the OA community can create and support simple factual lists about open access to science and scholarship)[3], provided by the School of Library and Information Science at Simmons College. From this list we retrieved details of 235 journals that flipped to OA prior to 2017[4]. Of these journals, 171 could be matched to journals indexed in the Web of Science (WoS) via matching of names and ISSNs.

The effect of flipping to OA will be measured by a descriptive analysis of the number of articles published by the flipped journals, the number of citations to those articles (using the matching citation algorithm implemented by the German Competence Centre Bibliometrics: http://www.forschungsinfo.de/Bibliometrie/en/index.php) and the respective journal IFs. Additionally, we will give an overview on when the most journals converted to OA.

We calculated the two years IF following the same methodology employed by Clarivate Analytics[5] for each of the 171 journals for each year. Based on this definition, the IF is defined as all citations to the journal in the current year to items published in the previous two years, divided by the total number of scholarly items (these comprise articles, reviews, and proceedings papers) published in the journal in the previous two years. In order to be able to compare IFs between different fields, they were field-normalized using the rescaling method introduced by Radicchi, Fortunato & Castellano (2008), in which the citation rate calculated in the IF definition is rescaled by dividing by the arithmetic mean of the citation rate of all articles in its discipline. The 252 subject categories included in WoS were applied for the field-normalization. In this classification system, journals can have more than one category, therefore we considered the mean citation rate of all articles in all categories of which this journal belongs to.

To study the evolution of citations at the article level, we calculate a relative citation (RC) count for each individual paper published within the journals in our dataset, normalised to account for different citation patterns across fields. The RC of a paper is calculated for each year by computing the sum of citations gained by the individual paper, divided by the average number of citations of all papers across its field(s) published in the same year. The window to compute the number of citations is three years (to be sure the articles receive their citation peak). An RC value above 1 means that a paper is cited more frequently than the average citation level for all papers in that field, and vice versa. To compare the citations for two groups of papers, we therefore calculate the mean RC of all papers for each group, referred to as the average of relative citations (ARC). In this way we can compare the ARC of multiple groups of papers across different years and different fields.

---

[2] http://www.science-metrix.com/?q=en/expertise/bibliometrics/methods
[3] http://oad.simmons.edu/oadwiki/Journals_that_converted_from_TA_to_OA
[4] See the list of journals here: https://github.com/momenifi/OASE
[5] https://clarivate.com/essays/impact-factor/metric

To be able to compare changes between our dataset of OA journals and a dataset of CA journals (obtained from the ScienceDirect website[6]) across multiple scientific domains we reclassified our dataset of journals into the six major domains of the Science Metrix classification[7] system described by Archambault et al. (2011). This classification system allows us to make coarser comparisons between major scientific domains in comparison to than the 252 subject categories in WoS.

Journals that had no flipping date indicated in the Open Access Directory were excluded from our analysis. Hence, we present results with a number of journals (n=171) that will be indicated in the Figure 1. Also, for our analyses of article published volumes, ARC and IF, we excluded the journals that had no articles, ARC and IF for at least one year of the considered years.

**Results**

IFs are calculated based on citations earned by articles published in the two past years, thus we expect to observe the impact of converting to OA at least one year after the flip. Due to the journal review time – e.g. when a journal flips, newly submitted articles will take several months to proceed through the review process. Therefore we considered the following year as the flipping point for journals which were flipped in the fourth quarter to ensure that articles reflect the OA model under which they were submitted. Figure 1 shows the distribution of years in which journals in our WoS dataset flipped. We observe a peak in the number of flipped journals in 2006, as well as a long-term steady increase in the number of journals that have converted to OA across all years. The peak in 2006 is caused by a large number of journal conversions carried out by two major publishers: Hindawi and the Spanish National Research Council. According to the Open Access Directory, no journals flipped in 2016.

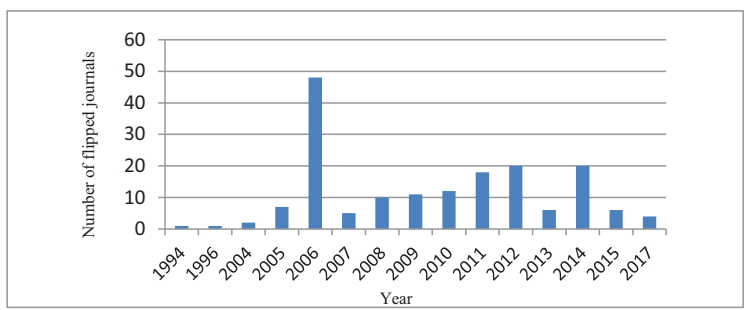

**Figure 1. Distribution of flipping years for 171 WoS journals.**

After a journal flips to OA, the costs of publishing transform from the reader side (i.e. subscriptions) and must be recovered from alternative sources, commonly in the form of

---

[6] https://www.sciencedirect.com/browse/journals-and-books?accessType=subscribed
[7] http://science-metrix.com/?q=en/classification

Articles Processing Charges (APC) paid by funders, institutes or authors directly (the 'Gold' OA model). Authors or institutes without access to sufficient funding sources may therefore be unable to publish in converted OA journals, which may influence the number of submitted articles. Conversely, some journals may flip to OA with the financial support of a funding body or society and not require publishing fees on the side of the author (the 'Diamond' OA model), which may influence published article volumes differently. Figure 2 shows the average number of articles published by journals four years before and after flipping. After excluding the journals that have at least one year with zero published articles and flipped after 2012, we obtain a dataset of 58 journals. From this figure, we observe a rising trend in the average number of publishing articles in the years preceding flipping when considering all journals (thick red line), followed by a decline in the number of published articles in the year after flipping (from an average of 142 published articles one year before flipping, to 132 one year after flipping). Two years subsequent to the journal flipping, published article volumes become more stable. When considering individual flipping years, we also observe a decrease in the number of articles published after flipping, with the sole exception of journals that flipped in 2009. Overall, these findings agree with those from Björk (2012) and others, that hybrid journals publish more articles in CA than OA.

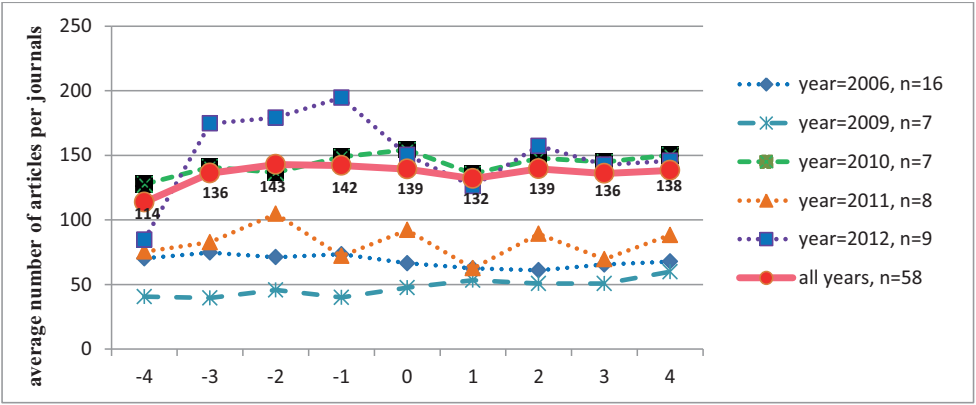

**Figure 2. Yearly average number of articles for flipped journals four years before and after flipping (n=number of journals and year is the flipping time). Read 0 on the x-axis as the year of flipping; -1 is the year before and 1 the year after flipping.**

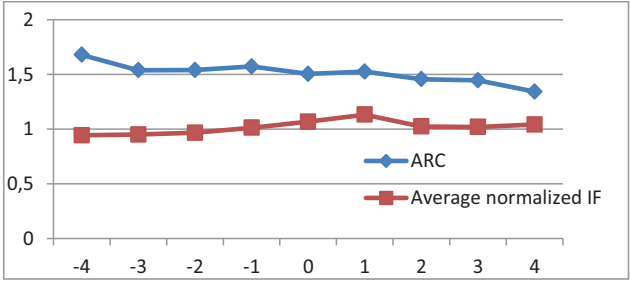

**Figure 3. Comparing the ARC of articles and average normalized IF of 43 flipped journals before and after flipping.**

Figure 3 shows the IF and ARC for journals and articles, respectively, in our dataset for the four years before and after flipping. We only included journals with IF and at least one article with and RC larger than zero for each of the years analysed and flipped until 2012. We observe a decrease in ARC after flipping, indicating a lower impact for articles published in these journals after flipping. However, the decrease in ARC occurs at the same time as a peak for IF one year after flipping. Our interpretation for this peak is that when the journals flip they probably also make the old articles OA, and so the increase in IF would be due to increased citations to articles that were published under the CA model, but have now been made OA. In general, we observe higher IFs in the years after flipping than before.

Flipping to OA can affect ARC and IF differently across different fields. To see the difference, we categorised the journals based on flipping years and fields, and compared their ARC and normalized IF with CA journals in the same field. Figure 4 shows a sample of the fields with the highest numbers of journals in each category. We observe that flipped journals experience more strongly fluctuating long-term trends of ARC and IF across the time period of our measurement, in comparison to those of the CA journals. However, for Health Sciences (A and B) we see an improvement in the ARC and IF after flipping, but journals in Applied Sciences (C and D) show no long-term advantage for these indicators for both years 2009 and 2010. However, we note that there are limited numbers of flipped journals across our samples in this analysis, and thus analysing these relationships more closely will be part of the focus of our future.

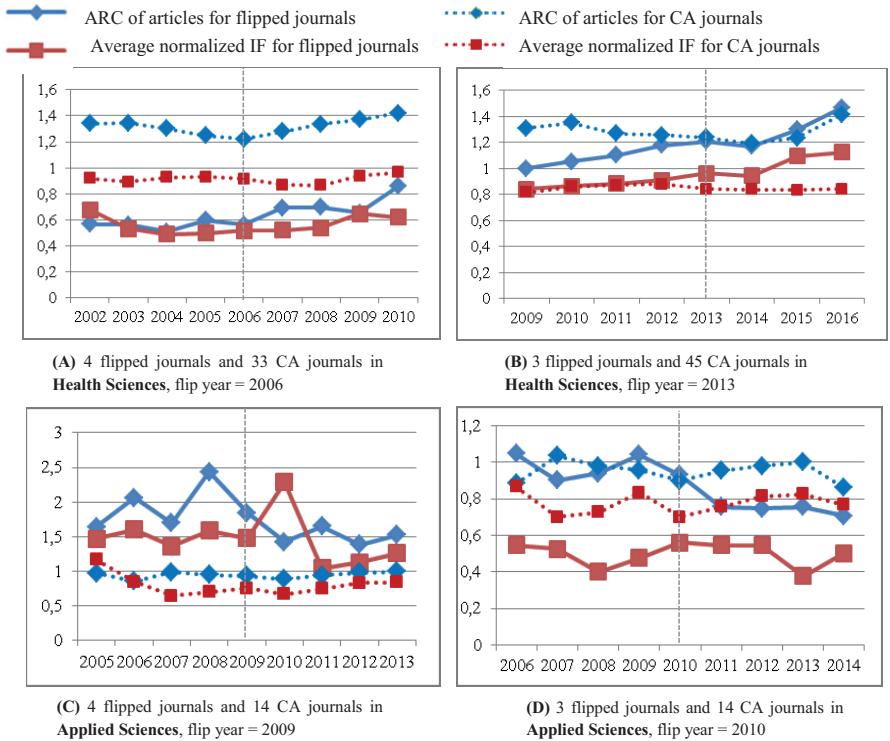

**Figure 4. Comparing ARC of articles and normalized IF of flipped and CA journals across some fields.**

**Conclusion and future work**

The literature reporting studies on flipped journals shows that journals IF usually increase after flipping (Busch, 2014a). Our results agree with these previous findings, but show that whilst IF increases in the years immediately following flipping, the article-level ARC decreases, in line with results found by Piwowar et al. (2018) and Archambault et al. (2014) that gold OA articles are cited less than CA articles. We also observe that a lower number of articles are published after flipping, pointing to a lower tendency amongst authors to submit to OA journals. So far, this work in progress studied only a limited number of flipped journals in the WoS with different flipping years and in different fields. Hence, future work will expand this study to Scopus to consider a higher number of flipped journals, and to research in greater detail the publication output of those journals before and after flipping, e.g. in terms of article numbers or length. We will also study in more detail the so-called "quality effect", which assumes that high quality publications profit proportionally more from OA, as they are more citable than low-quality publications (Ottaviani, 2016). The question we seek to answer is whether journals with high IFs benefit significantly more from flipping to OA than journals with medium or low IFs. Our bibliometric study will be complemented with more qualitative information from interviews with authors, editors, and publishers to reveal their attitudes towards journal flipping and OA, their expectations regarding journal quality and indicators as well as their motivation to change the publication model. With this we hope to better show such interwoven factors that influence bibliometric indicators of flipped journals.

**Acknowledgement**

This work is supported by BMBF project OASE, grant number 01PU17005A.